# Dynamic Canonical and Microcanonical Transition Matrix Analyses of Critical Behavior

David Yevick and Yong Hwan Lee

Department of Physics

University of Waterloo

Waterloo, ON N2L 3G7

**Abstract:** By monitoring the sampling of states with different magnetizations in transition matrix procedures a family of accurate and easily implemented techniques are constructed that automatically control the variation of the temperature or energy as the calculation proceeds. The accuracy of the method for a single Markov chain exceeds that of standard transition matrix procedures that accumulate elements from multiple chains.

**Introduction:** Large physical systems are generally characterized by the statistical behavior of one or more (typically macroscopic) variables $\vec{E}(\vec{\alpha})$ that depend on a large number of stochastically varying (microscopic) parameters, $\vec{\alpha}$. Numerous methods have accordingly been developed to preferentially sample either broad or physically important low probability regions of the microscopic configuration space including importance sampling [1], multicanonical [2] [3] [4] [5] [6] [7] [8] [9] [10] [11] [12] and Wang-Landau [13] [14] [15] methods.

The transition matrix formalism [16] [17] [18] [19] [20] [21] can employ any of these procedures to construct a matrix $\mathbf{T}$ such that $T_{ij}$ corresponds to the probability that a Markov chain state in a histogram bin $E_i$ transitions to bin $E_j$ after the displacement $\delta\vec{\alpha}$. The normalized eigenvector of $\mathbf{T}$ with unit eigenvalue, which is easily obtained by repeatedly multiplying a random vector by $\mathbf{T}$ then coincides with the desired probability distribution $p(E)$. [22]

A unique feature of the transition matrix procedure is that since all accepted and rejected transitions are recorded, effectively arbitrary acceptance rules that do not necessarily preserve detailed balance can be employed. These include rules based on the multicanonical or Wang-Landau acceptance probabilities [23] [24] [25] [26] [27], the transition probabilities between microscopic states [16] [17], the ratio of transition matrix elements [28] [29] [17], and the exclusion of transitions to bins that have been previously visited a larger number of times. [24] [22] [30] A renormalization procedure has further been advanced that accelerates the analysis of large systems by convolving the results of transition matrix calculations of smaller systems. [31] [32] [33]

Despite the apparent generality of the transition matrix method, the accuracy of calculations based on different biased sampling methods varies widely. For example, a comparison of several approaches to the calculation of the specific heat in the Ising model [34] [35] [32] demonstrated that procedures based on accumulating transition matrix elements in calculations based on canonical, Metropolis sampling [36] at temperatures that varied gradually from a large initial value to zero as the calculation progresses were considerably more accurate than quasi-microcanonical methods associated with restricting the calculation to a narrow energy region that was similarly steadily displaced from high to low energies. Further improvements in precision were achieved by controlling the temperature according to a schedule that increased the number of Markov steps for temperatures in the Metropolis acceptance rule near the critical temperature and by collecting the transition matrix elements from numerous independent computations in a single transition matrix.

Recently the origin of numerical error in the transition matrix procedure was described in terms of the evolution of the Markov chain in two-dimensional energy-magnetization ($E-H$) space. [32]  A numerical study established that the diffusion velocity of the Markov chain for Metropolis sampling in this parameter space was nearly constant. However, when the temperature in the Metropolis rule is close to the critical temperature, the area of the $E-H$ plane that must be sampled by the Markov chain expands greatly. The number of Markov steps required for the chain to diffuse a distance equal to several times the extent of this region was then found to be on the order of the number that insured reasonable numerical accuracy in specific heat computations.

In this paper, the results of the previous paragraph are further exploited, to our knowledge for the first time, by demonstrating that if at a given temperature or energy the number of Markov chain realizations is sufficiently large to insure a stable distribution of samples over the physically accessible $E-H$ region, the accessible phase space is sufficiently sampled. Consequently, by monitoring the distribution of samples in the $E-H$ plane several simple, easily programmed adaptive techniques are constructed that automatically generate a temperature schedule based on the expansion and contraction of the accessible phase space with temperature or energy. A high level of accuracy can accordingly be attained without prior knowledge of the system behavior near the critical point. A novel justification for the technique is further provided based on the variation in the ratios of transition matrix elements with magnetization at a fixed energy.

While the two-dimensional $32\times 32$ Ising model is employed below as a benchmark to establish the accuracy of the adaptive method relative to previous results [32] [18] [28] [31] [34], the procedure can be adapted to any calculation involving a phase transition if the order parameter is substituted for the magnetization. The accuracy afforded by the algorithm would depend however on the functional dependence of the transition matrix elements on the order parameter at fixed energy as well as the area of the accessible $E-H$ space near the critical point and therefore is not expected to be easily predictable in advance.

**Computational procedure:** Two classes of transition matrix calculations are distinguished below. In the *canonical* method the transition matrix elements are accumulated subject to the Metropolis acceptance rule while the temperature is either increased or decreased between infinity and a suitably small value, typically close to one-half the critical temperature.

*Microcanonical* procedures instead implement an acceptance rule that rejects transitions of the Markov chain to configurations with energies outside an interval of width $\Delta E_{\text{width, microcanonical}}$ centered at an energy $\bar{E}_{\text{microcanonical}}$. The value of $\bar{E}_{\text{microcanonical}}$ is then is displaced between the energy limits of the system as the calculation proceeds. (Techniques that rapidly adjust the energy range as the calculation proceeds but still reject transitions outside the range can be termed quasi-microcanonical.)

The adaptive procedure is implemented as follows. The temperature is first fixed at its upper or lower limit. Then, for every realization of the Markov chain to a state with a magnetization $k$, the $k$ :th element,

$H_k$, of a histogram $H$ is updated at each Markov step. A second histogram, $H_{old}$, is only updated after each $N_{magnetization}$ steps of the Markov chain by setting $H_{old} = H$. Immediately before each update, the mean value of the absolute values of the elements in the array

$$H - H_{old} \left( \frac{\sum_k H_k}{\sum_k H_{old,k}} \right) \tag{1}$$

Is compared to a threshold value $H_{threshold} N_{magnetization}$. If the mean value is less than the threshold, the temperature is changed by an amount $\pm\Delta(1/T)$ (or in microcanonical methods the energy is changed by $\pm\Delta\bar{E}_{microcanonical}$) and both $H$ and $H_{old}$ are reset to zero. The value of $H_{threshold}$ is typically set empirically by examining the error for a few test runs in the same manner that the number of steps is determined in a standard transition matrix calculation. To guard against the program from running for an excessively long time in the event that the convergence criterion cannot be efficiently satisfied, the temperature is also updated if the average value of the elements of $H$ that have been visited more than $N_{minimum}$ times exceeds a limiting value $H_{limit}$.

**Numerical results:** The precision of the adaptive transition matrix procedure will now be established by evaluating the specific heat of the two dimensional Ising model with zero external magnetic field, periodic boundary conditions, and a unit amplitude ferromagnetic interaction between an array of 32x32 integral spins. The numerical parameters are $N_{minimum} = 4 \times 10^3$, $H_{limit} = 3 \times 10^4$, $N_{magnetization} = 1 \times 10^5$ and $H_{threshold} = 8 \times 10^{-4}$. Additionally the inverse temperature shift $|\Delta(1/T)| = 0.00075$, while $\Delta E_{width, microcanonical} = 24$ and $|\Delta\bar{E}_{microcanonical}| = 8$ in the microcanonical calculations. The precision of the calculations is enhanced by rejecting transitions to states with a negative magnetization.

To illustrate the procedure, Figure 1 first displays the number of steps employed in a representative canonical calculation as a function of normalized inverse temperature for inverse temperatures varying from $0.65$ to $0$ in steps of $\Delta(1/T) = -0.00075$. The number of steps required to fulfill the convergence criterion is, as expected, small except near the critical temperature. The infrequent large fluctuations in the figure are most likely generated by the limiting criterion which can prematurely terminate the calculation if the number of visits to a small number of highly populated magnetization bins rapidly exceeds $H_{limit}$.

The origin of the marked increase in the number of samples near $T_c$ in Figure 1 is apparent from Figure 2, which displays $H$ just before updating the temperature, normalized such that the maximum of each curve is unity, for a canonical calculation with inverse temperatures varying from $0.65$ (the rightmost curve) to $0$ (the leftmost curve) in steps of $\Delta(1/T) = -0.0375$. For low temperatures the system is nearly completely magnetized and the histograms are therefore narrow while for high temperatures the histograms resemble Gaussian functions. In both cases, a steady state distribution is obtained after a small number of realizations. In contrast, close to the critical temperature as discussed in detail in [35] [32], the accessible range of magnetizations is broad. Numerous transitions are therefore required for the Markov chain to diffuse repeatedly throughout the physically accessible energy-magnetization region.

To analyze the relationship between the computational accuracy and the requirement that the number of realizations at a given temperature value be sufficient to insure that the distribution of Markov chain visits among states of different magnetization be approximately invariant, the ratios of the $\Delta E = 4$ to the $\Delta E = 8$ transition matrix elements are displayed as a function of magnetization for an energy of $-1384$ in Figure 3. While statistical fluctuations are substantial, this ratio clearly decreases with magnetization, an effect which becomes far more pronounced away from the critical energy. Consequently, unless the distributions of Figure 2 are close to the exact values, the transition matrix elements will be incorrectly evaluated.

Since temperature averages are employed to determine the specific heat, it is most accurately calculated with the canonical procedure. Accordingly, Figure 4 overlays the results of twenty separate calculations as a function of normalized inverse temperature. In each of these calculations, the transition matrix was constructed according to the procedure of the preceding section as the inverse temperature first evolved from $0.75$ to $0$ and then back to $0.75$ with steps of magnitude $|\Delta(1/T)| = 0.00075$. The histograms of the total number of realizations employed in each calculation and of the maximum values of the specific heat curves are presented respectively in Figure 5 and Figure 6. The histogram width is similar to that of Fig. 4 of [32], which employed a heuristic profile to specify the number of required steps as a function of temperature and further accumulated the elements of the transition matrix from $60$ independent Markov chains with $6 \times 10^7$ steps for each realization. [37] [38] [35] These results clearly indicate that the precision of the adaptive method exceeds that of standard procedures for an equivalent number of realizations.

While microcanonical and quasi-microcanonical calculations of the specific heat are intrinsically less accurate than those employing the canonical procedure, in other non-statistical mechanics contexts microcanonical methods can prove advantageous. [39] In Figure 7 the result of a typical microcanonical calculation in which the unnormalized value of $\bar{E}_{\text{microcanonical}}$ is first lowered from $0$ to $-2056$ (lower solid curve) is compared to that of a similar calculation in which $\bar{E}_{\text{microcanonical}}$ is instead raised from $-2056$ to $0$ (upper solid curve) and to the exact result [40] (dashed curve). The error of the two curves, which is smaller when the initial state is magnetized, is systematic but can be reduced by accumulating the transition matrix elements from both of the calculations. Applying this procedure to the calculation of Figure 7 and repeating this procedure $20$ times yields the histogram for the specific heat maxima of Figure 8 where $\approx 7 \times 10^8$ realizations are employed in each calculation. While slightly less accurate than, Fig. 2 of [32], which however employed $60$ independent Markov chains and a factor of $5$ larger total number of realizations, further improvements can be realized with a technique somewhat similar to the bidirectional procedure of Ref. [35]. Namely, as $\bar{E}_{\text{microcanonical}}$ is lowered from $0$ to $-2056$, $\Delta \bar{E}_{\text{microcanonical}}$ is set to $8$ instead of $-8$ after every third set of $N_{\text{magnetization}}$ steps. The overlaid curves of twenty independent evaluations of the specific heat are displayed in Figure 9, and the corresponding histogram for the specific heat maxima is presented in Figure 10.

**Discussion and Conclusions:** The adaptive transition matrix procedure possesses clear advantages over transition matrix methods that accumulate transition matrix elements from multiple independent Markov chains that evolve in temperature and energy according to a specified schedule. In particular, the technique of this paper is not only simpler to program, but also automatically generates an appropriate schedule by dynamically monitoring the evolution of the Markov chain as a function of magnetization, which is shown above to be directly related to the accuracy of the elements of the transition matrix. However, the techniques of this paper depend on a small number of parameters which can only be fully optimized after considerable additional effort. Additionally, the criterion for effecting a temperature change is not unique. The results above however establish the validity of the general class of such methods and suggest directions for future studies. For example, the accuracy of the microcanonical approach was enhanced by requiring the Markov chain to propagate intermittently in the direction of higher temperature as the temperature was lowered. A further investigation of the precise manner in which this procedure should be carried out and whether it can be more fully integrated, for example, with the bidirectional technique of [35] [32] could yield further improvements as well as a deeper understanding of the exact origin of the numerical error in transition matrix procedures. The numerical error of the transition matrix method can as well presumably be further reduced by combining multiple independent Markov chains with adaptive sampling. Finally, an extension of the algorithm to non-magnetic phase transitions in which the relevant order parameter would be substituted for the magnetization in the criterion that determines the threshold for an energy or temperature change would also be of considerable interest.

**Acknowledgements:** The Natural Sciences and Engineering Research Council of Canada (NSERC) is acknowledged for financial support.

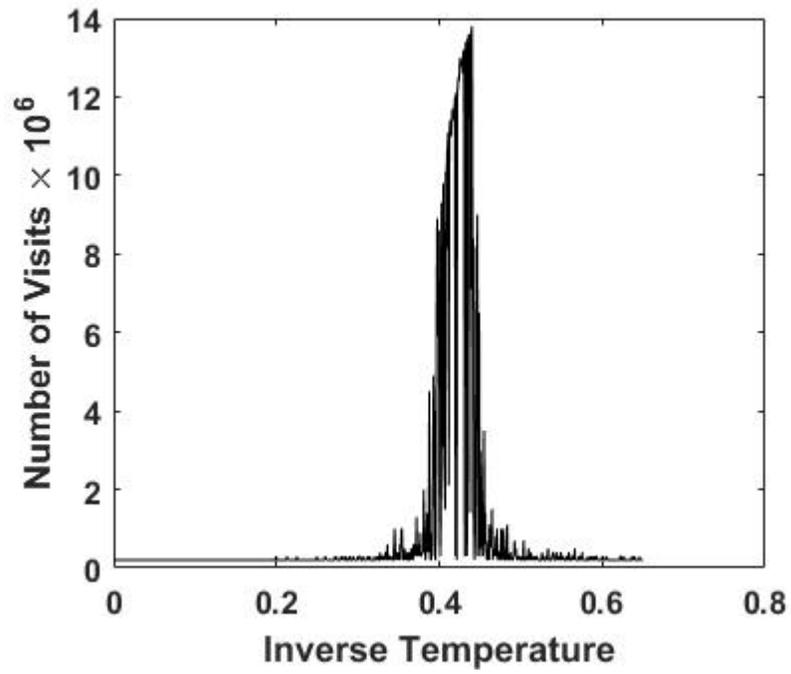

*Figure 1: The number of steps taken in the canonical calculation as a function of inverse normalized temperature for inverse temperatures varying from 2.6 to 0 in steps of 0.003*

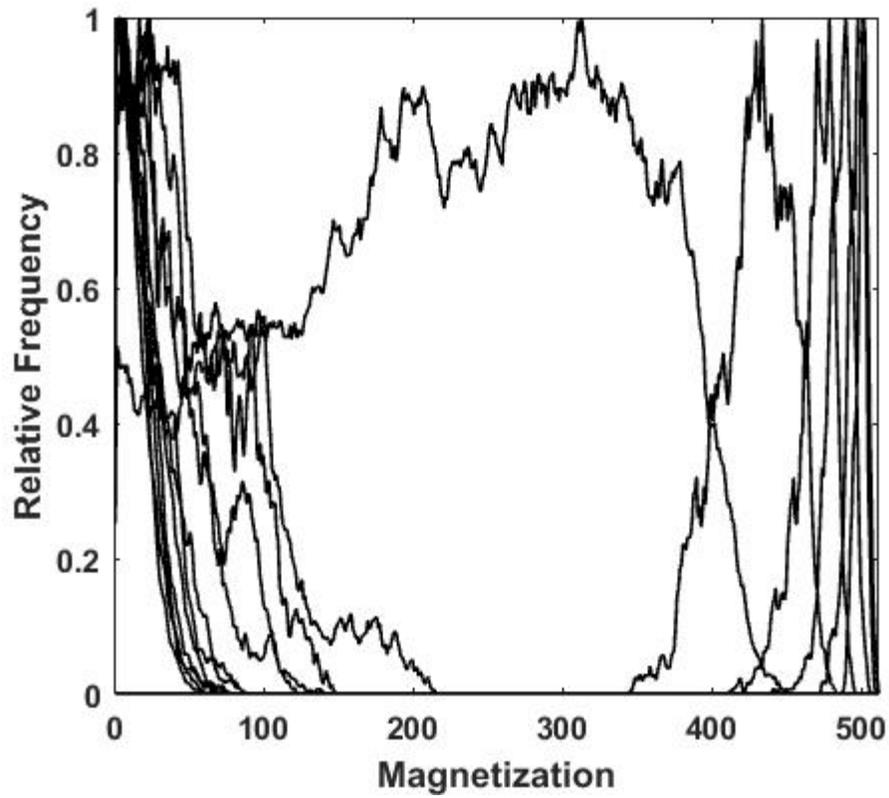

Figure 2: The distribution of visits of the Markov chain as a function of magnetization for inverse temperatures varying from 2.6 (the rightmost curve) to 0 (the leftmost curve) in steps of 0.15

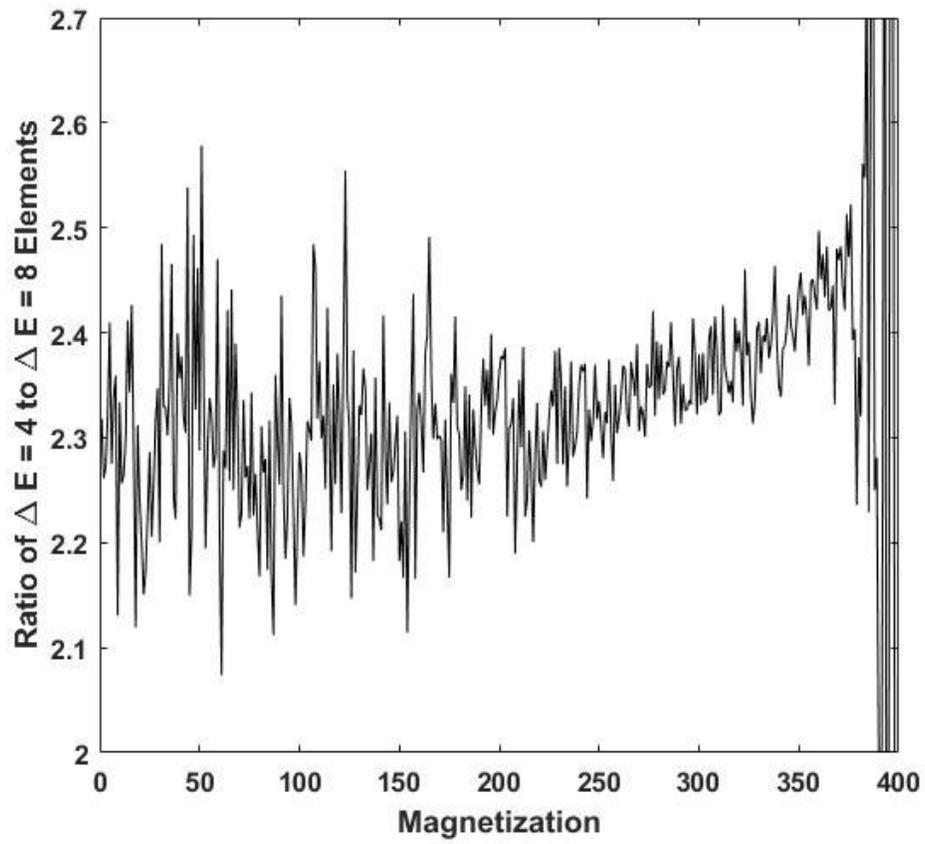

Figure 3: The ratio of the $\Delta E = 4$ to the $\Delta E = 8$ transition matrix elements as a function of magnetization for an energy of -1384.

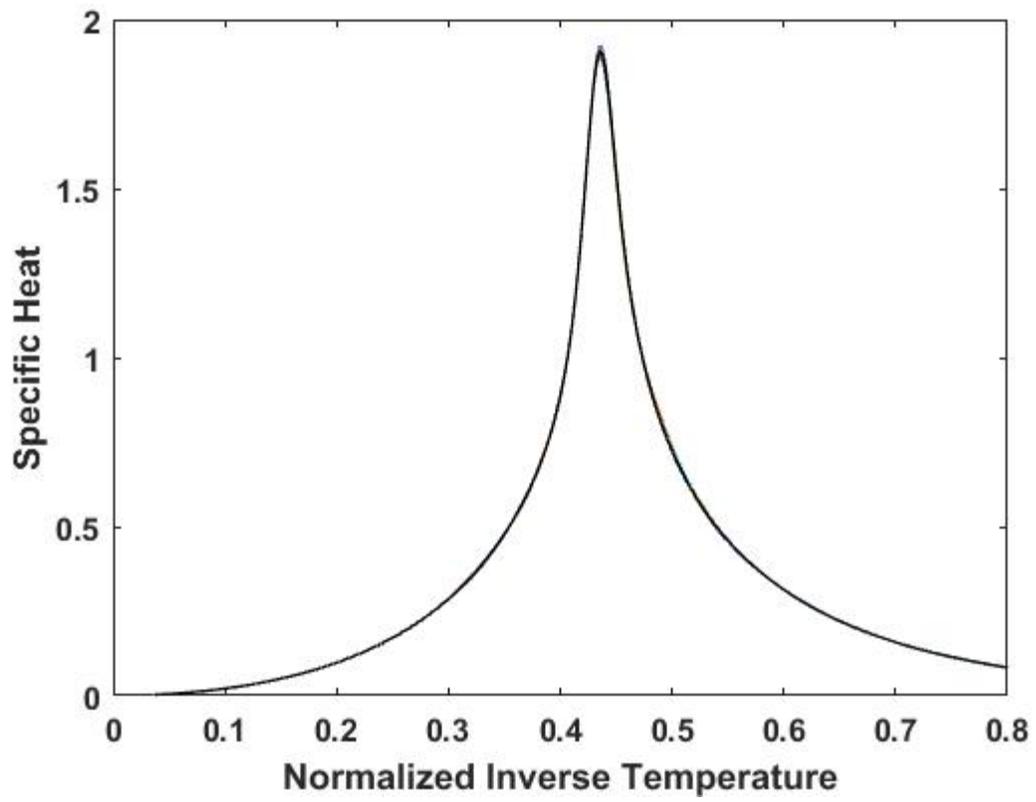

Figure 4: The superimposed results of twenty separate canonical calculations of the specific heat as a function of normalized inverse temperature

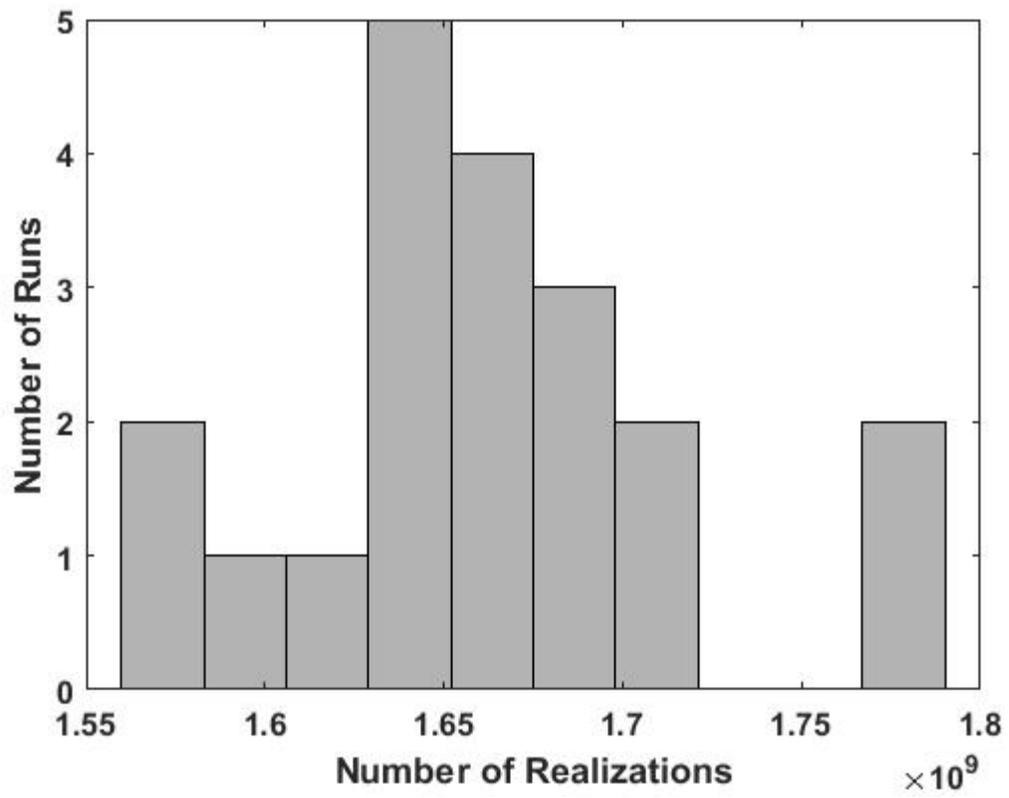

Figure 5: Histograms of the total number of Markov steps employed in the twenty calculations of the preceding figure

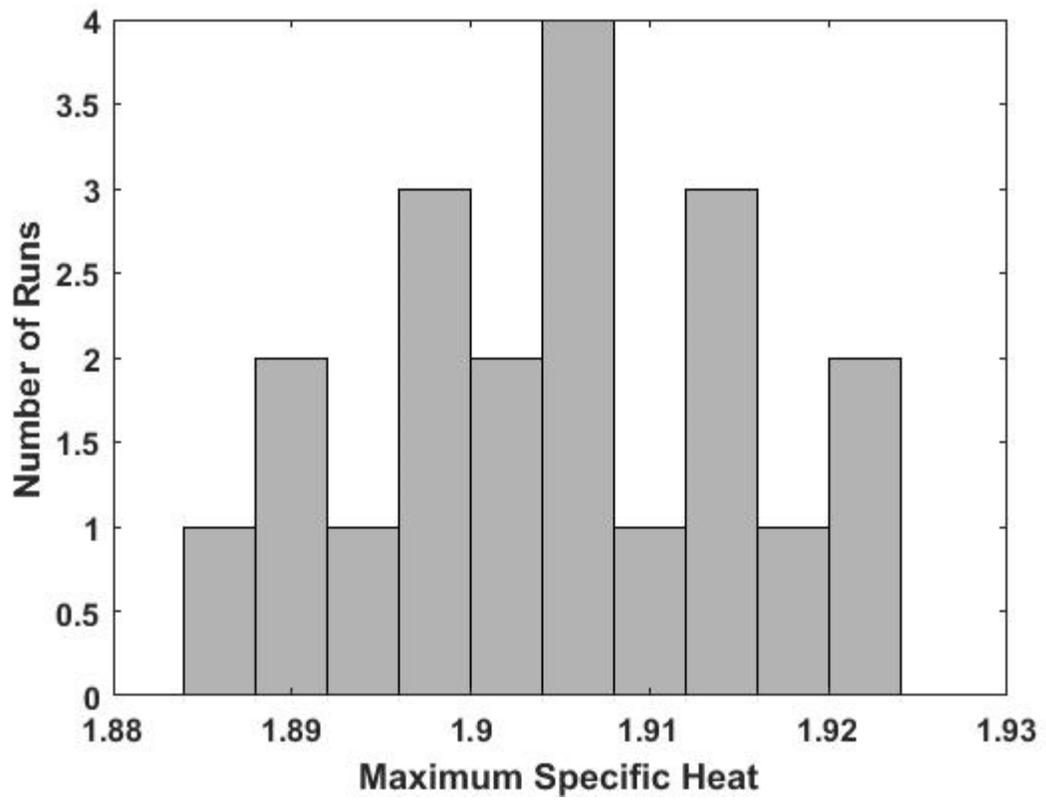

Figure 6: As in Fig. (4) but for the maximum of the specific heat curves

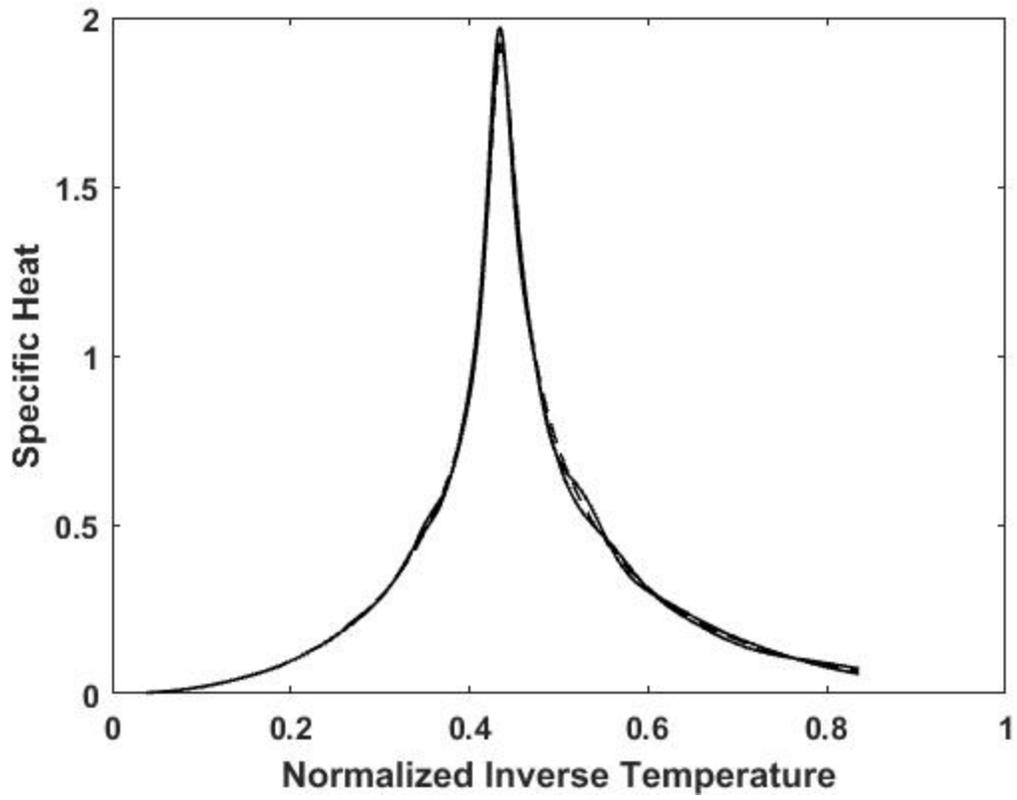

Figure 7: The specific heat generated by microcanonical calculation in which the inverse temperature is lowered from 3 to 0 (lower solid curve), raised from 0 to 3 (upper solid curve) and the exact result (dashed curve).

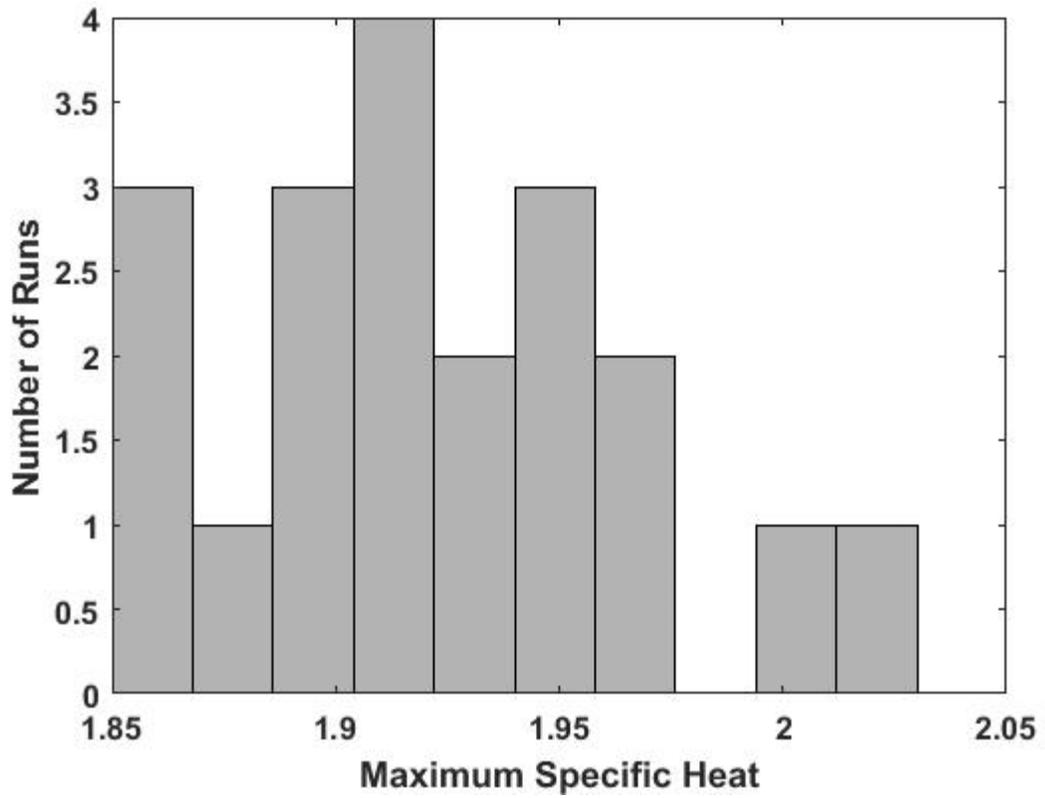

*Figure 8: The histogram of the maxima of the specific heats obtained after 20 independent calculations when both temperature sweeps of the preceding figure are employed to populate the same transition matrix*

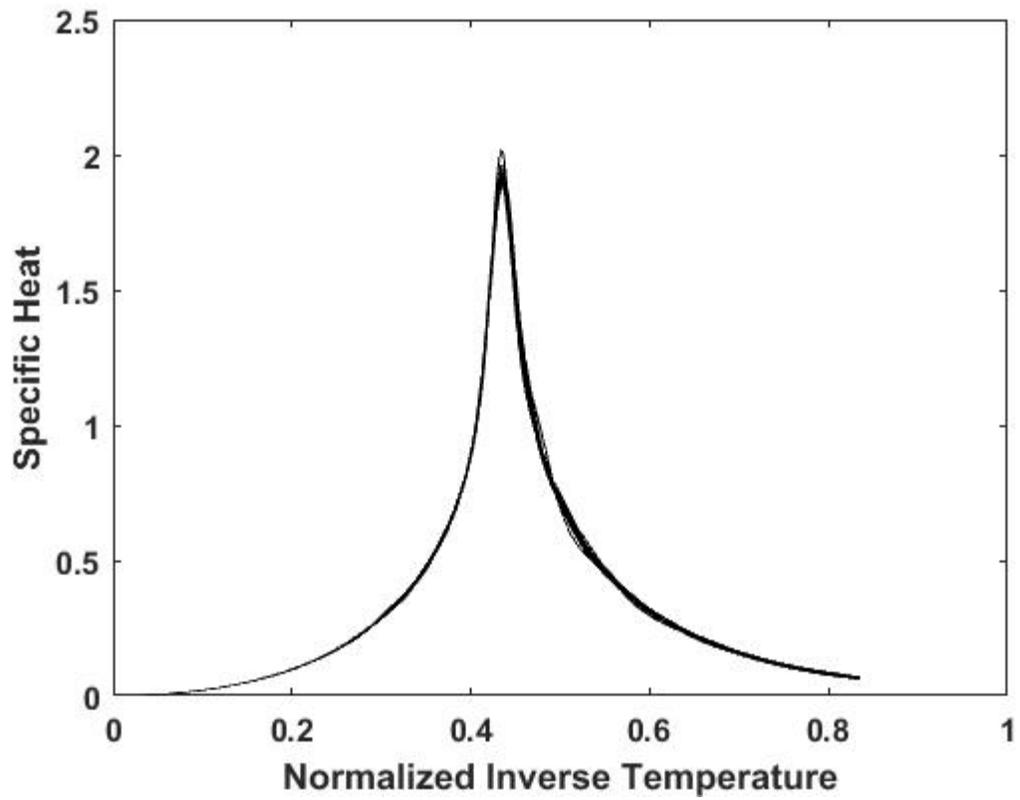

*Figure 9: Twenty independent microcanonical calculations of the specific heat in which while the inverse temperature is lowered from 3 to 0, the temperature is raised instead of lowered **at** every third temperature change.*

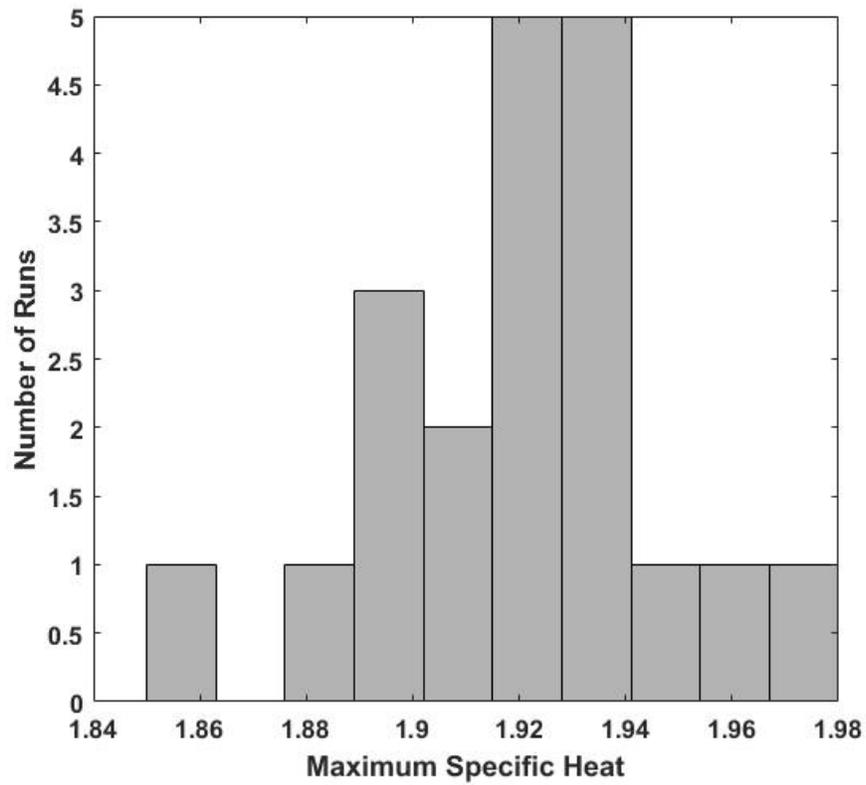

*Figure 10: The histogram of the maxima specific heat for the calculations of the previous figure in which the temperature is raised instead of lowered **at** every third temperature change.*